\def\year{2015}
\documentclass[letterpaper]{article}
\usepackage[utf8]{inputenc}
\usepackage{aaai}
\usepackage{times}
\usepackage{helvet}
\usepackage{courier}
\usepackage{graphicx}
\usepackage{subfigure}
\usepackage{units}
\usepackage{amsmath}
\begin{document}

\title{Voting Behaviour and Power in Online Democracy:\\A Study of LiquidFeedback in Germany's Pirate Party}

\author{
  Christoph Carl Kling\textsuperscript{1}\ 
  Jérôme Kunegis\textsuperscript{1}\ 
  Heinrich Hartmann\textsuperscript{1}\ 
  Markus Strohmaier\textsuperscript{1,2}\ 
  Steffen Staab\textsuperscript{1}
 \\
  \textsuperscript{1} University of Koblenz--Landau, Rhabanusstraße 3 , 55118 Mainz, Germany \\
  \textsuperscript{2} GESIS -- Leibniz Institute for the Social Sciences, Unter Sachsenhausen 6--8, 50667 Cologne, Germany \\
  \{ckling,kunegis,hartmann,strohmaier,staab\}@uni-koblenz.de 
}
\maketitle

\begin{abstract}
In recent years, political parties have adopted \emph{Online Delegative Democracy} platforms such as LiquidFeedback to organise themselves and their political agendas via a grassroots approach. A common objection against the use of these platforms is the delegation system, where a user can delegate his vote to another user, giving rise to so-called super-voters, i.e.\ powerful users who receive many delegations.
It has been asserted in the past that the presence of these super-voters undermines the democratic process, and therefore delegative democracy should be avoided.
In this paper, we look at the emergence of super-voters in the largest delegative online democracy platform worldwide, operated by Germany's Pirate Party.
We investigate the distribution of power within the party systematically, study whether super-voters exist, and explore the influence they have on the outcome of votings conducted online.
While we find that the theoretical power of super-voters is indeed high, we also observe that they use their power wisely.
Super-voters do not fully act on their power to change the outcome of votes, but they vote in favour of proposals with the majority of voters in many cases thereby exhibiting a stabilising effect on the system. We use these findings to present a novel class of power indices that considers observed voting biases and gives significantly better predictions than state-of-the-art measures.

\end{abstract}

\section{Introduction}
In the last decade, the World Wide Web has increasingly been adopted for facilitating political processes and conversations \cite{lietz2014}. The Web has also sparked the development of novel voting and democracy platforms impacting both societal and political processes. Today, a wide range of online voting platforms are available, based on different democratic methods such as consensual decision making, liquid democracy \cite{paulin2014through} or dynamically distributed democracy \cite{tenorio2014towards}. These platforms are becoming increasingly popular and political movements and parties have started adopting them to open up and facilitate political coordination. In contrast to experimental data or simulations (e.g.\ from game theory), the behaviour of voters on these platform is realistic, i.e.\ voting takes place in a natural environment and the decisions of voters have a real political impact. Having such a natural setting is crucial for studying voting behaviour in real life political movements and for validating research on voting behaviour and measures of power~\cite{loewenstein1999experimental}. Yet, this kind of data has historically been elusive to researchers.

LiquidFeedback represents a popular platform which implements support for \textit{delegative democracy}. In contrast to a representative democracy, all voters in a delegative democracy in principle are equal. Each voter can delegate his vote to another voter, raising the voting weight of the delegate by one. The delegate again can delegate his voting weight to a third user and so forth, creating a transitive delegation chain. A key innovation of delegative democracy platforms is the ability of every voter to revoke his delegated votes at any point, preserving full control over his votes and allowing for the emergence of dynamic delegation structures in contrast to representative voting systems. Votes are public and pseudonymous, and therefore both individual and collective voting behaviour can be analysed. A common objection against the use of these platforms is the nature of delegations, as they can potentially give rise to so-called \textit{super-voters}, i.e., powerful users who receive many delegations. It has been asserted in the past that the presence of these \textit{super-voters} undermines the democratic process, and therefore delegative democracy should be avoided.

\begin{figure*}
  \centering
  \subfigure[Delegation network]{
  \includegraphics[width=0.6\textwidth]{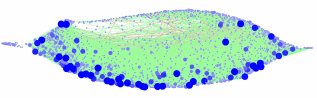}
  \label{fig:graphdrawing}
    }
  \centering
   \subfigure[User activity]{
  \includegraphics[width=0.35\textwidth]{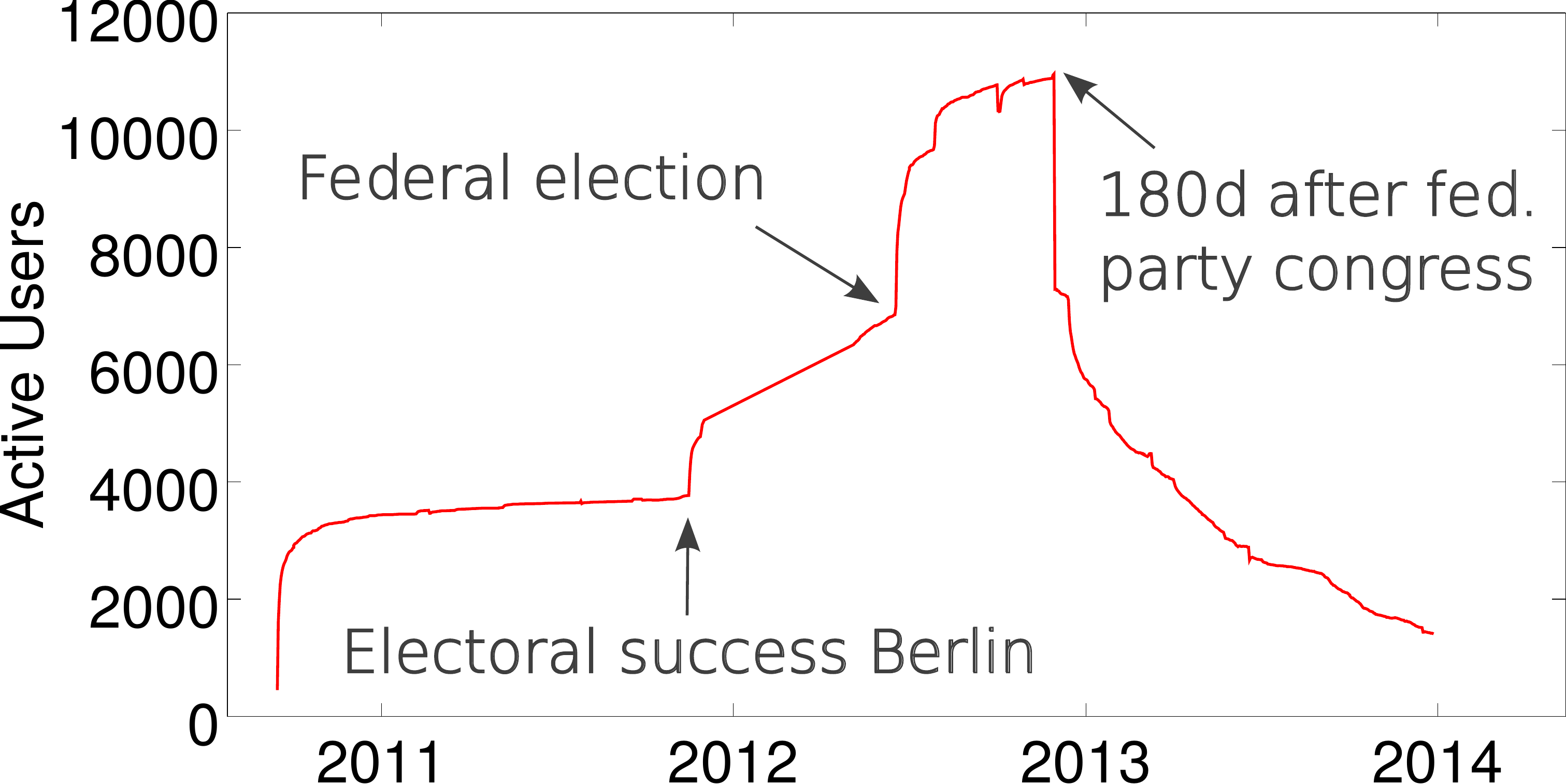}
   \label{fig:active_users}
  }
  \caption{
   \subref{fig:graphdrawing} The delegation network. A node denotes a user of LiquidFeedback (a Pirate Party member), edges denote delegations. Red edges denote removed delegations. Node size and color correspond to the number of delegations received by other users.  The layout was computed using the dominant eigenvectors of the network's stochastic adjacency matrix.
We can observe (i) the emergence of a large connected component of users and (ii) the existence of \textit{super-voters}, i.e.\ voters that have received a large number of delegations.
    \subref{fig:active_users}
    User activity. Active users on the LiquidFeedback platform of the German Pirate Party over time. Users are labelled inactive after 180 days without login. Several events led to a rise and decrease in activity.
  }
\end{figure*}

\smallskip
\noindent
\textbf{Problem.} In order to assess the true potential and limitations of delegative democracy platforms to facilitate political discourse and agenda setting, we first need to understand the behaviour of voters and \textit{super-voters}, and the power they wield.
Tapping into the complete voting  history and delegation network from world's largest delegative democracy platform (operated by the German Pirate Party), we want to understand \textbf{(i)~voting behaviour:} \emph{how people vote} in delegative democracy platforms such as LiquidFeedback, and \emph{how they delegate votes} to \textit{super-voters}. Based on these insights, we want to study the \textbf{(ii)~voting power:} \emph{how power can be assessed} in online democracy systems and \emph{how it is used}.

\smallskip
\noindent
\textbf{Approach.} We tackle these problems by analysing the voting behaviour of members of the German Pirate Party from 2009--2013. The German Pirate Party has adopted LiquidFeedback as their online delegative democracy platform of choice. We look at the delegation network of users over time and identify the emergence of power structures and \textit{super-voters} within the party. Next, we discuss and apply a series of established power indices from game theory and political science theory (specifically the Shapley and Banzhaf power index \cite{Shapley1954,Banzhaf1965}) to assess the theoretical power of super voters. Then, we compare the theoretical power of super voters with their potential as well as their exercised power based on real world voting data. Our analysis reveals \emph{a clear gap} between existing theoretical power indices and actual user voting behaviour. As a result, we develop and present a new class of power indices that better captures voting behaviour. We evaluate the proposed power indices with data from the LiquidFeedback platform.

\smallskip
\noindent
\textbf{Contributions.} This article makes two main contributions: The first one is \emph{empirical}: We provide unique insights into the evolution of voting behaviour and power in an emerging political movement: Germany's Pirate Party. 
Our analysis spans almost the entire life time of the party's online voting platform. 
We find that \emph{\textit{super-voters} exist}, that their \emph{theoretical and potential power is high}, but we also observe that they have \emph{a stabilising effect} on the voting system and that they \emph{use their power wisely}: \textit{Super-voters} do not fully act on their power to change the outcome of votes, and they vote in favour of proposals with the majority of voters in many cases. These findings represent the basis for our second contribution, which is \emph{methodological:} We investigate the potential and limitations of existing power indices from game theory and political science theory and identify \emph{a gap} between these models and what we observe in real-world data. Addressing this gap, we propose \emph{a novel class of power indices} that better captures voting behaviour. We evaluate the proposed indices using real voting data from the LiquidFeedback system. Via experiments, we demonstrate that the introduced power indices represent an improved way of characterizing the power of super voters in delegative democracy platforms.

\section{Related Work}
\label{sec:related}
\smallskip
\noindent
\textbf{Online Voting.}
Existing literature on online voting in the political setting is in part focused on
the design of secure and reliable voting mechanisms~\cite{conf/sp/KohnoSRW04},
applications for decision support~\cite{conf/hicss/RobertsonWP07} or
the evaluation of voting tools for public participation
to support traditional democratic systems~\cite{watkins2008survey}.
The availability of large-scale datasets on individual voting behaviour in non-political online communities enabled the exploration of latent mechanisms behind individual voting decisions. One interesting pattern is the rating bias found in online voting systems.  In \cite{journals/corr/abs-0909-0237}, Amazon, IMDB and BookCrossing show a systematic tendency towards positive (Amazon and IMDB) or negative (BookCrossing) ratings which approximately follows a beta distribution.
Another aspect of online voting systems is the impact of the similarity between voters and the voted candidates in elections of persons. For instance, in \cite{leskovec2010}, a positive impact of user similarity on the support of a candidate was found in the promotion process of Wikipedia administrators.
In \cite{muchnik2013social}, \textit{herding effects} were discovered in experimental settings, showing a significant impact of prior positive ratings of comments in an online community. Comments which already received a positive vote were more likely to receive a positive vote by other users.

All these findings were based on democratic voting systems with equal power distributions.
In this paper, we look, for the first time, at the aspect of power distributions in online voting systems with delegations.

\smallskip
\noindent
\textbf{Power Indices.} Research on the distribution of power in voting systems lead to the development of power indices. Power indices are numerical indicators designed to measure the ability of voters to influence voting outcomes.
The most common power indices are that of \citeauthor{Shapley1954} (\citeyear{Shapley1954}) and that of \citeauthor{Banzhaf1965} (\citeyear{Banzhaf1965}).
Both indices are based on game theory and are mostly popular due to their simplicity.
Other power indices try to capture the parliamentary reality, e.g.\ by limiting the index to majorities by minimal coalitions~\cite{deegan1978new,packel1980axiomated}.
As voting weights change frequently in delegative democracies, no fixed coalitions are formed and thus minimal coalitions are just as likely as any other coalition.

Straffin (\citeyear{Straffin1977}) gave a probabilistic formalisation of power indices, analysed underlying assumptions and gave recommendations when to apply which index based on subjective assessments of the assumptions.
Gelman et al.\ criticised the simplicity of the game-theoretic approaches by suggesting an Ising model for modelling dependencies between voters, e.g.\ common administrative regions \cite{Gelman.Katz.ea2002MathematicsandStatistics}. However, that study lacks the appropriate data for fitting the model as it relies on aggregated voting results and therefore cannot consider decisions at the individual level.
In this paper, we show how to utilise user-based voting behaviour to derive adjusted power indices and conduct the first objective evaluation of power indices on large real-world voting data with constantly changing voting weights.

\smallskip
\noindent
\textbf{Delegative Democracy.} First steps towards the direction of a delegative democracy were published in 1884 by Charles L.\ Dogson, better known under his pseudonym Lewis Carroll.
In his book about the mathematical properties of voting mechanisms, he proposes a voting scheme where elected candidates may delegate their votes to other candidates. The delegated votes then can be further passed to other candidates~\cite{dodgson1884principles}. A review of further works which influenced the development of the concept of a delegative democracy can be found in~\cite{jabbusch2011liquid,paulin2014through}. Based on these ideas, the novel concept of delegative voting was developed and recently popularised. A formalisation of a delegative democratic system is given in~\cite{yamakawa2007toward}.
The implementation of delegative voting systems is non-trivial as loops in the delegation network have to be detected and resolved and regaining votes potentially can affect a long delegation chain.

\smallskip
\noindent
\textbf{Democracy Platforms.}
Existing software implementations of delegative democracy include
\textit{LiquidFeedback (\mbox{liquidfeedback.org})},
\textit{Agora Voting (\mbox{agoravoting.com})},
\textit{Get\-Opinionated (github.com/getopinionated)} and
\textit{Democracy OS (democracyos.org)}.
In this paper, we study the online voting platform of the German Pirate
Party, which is based on LiquidFeedback,
a free software that implements an online platform in which votes can be
conducted, and users can delegate their vote to other users.  LiquidFeedback
was adopted by the German Pirate Party in
May 2010~\cite{paulin2014through} and has 13,836 users as of January 2015.


\smallskip
\noindent
\textbf{Pirate Parties.}
Pirate parties are an international political movement with
roots in Sweden \cite{fredriksson2013open}, where legal cases related to copyright
law led to the formation of a party advocating modern copyright
laws and free access to information \cite{miegel2008pirates}. The scope of the
party quickly broadened and nowadays active pirate parties
exist in 42 countries. The German Pirate Party is the largest
of all pirate parties with 24,438 members as of January 2015.

\begin{table}[b]
\center

\caption{\textbf{The LiquidFeedback dataset.} Obtained from the delegative democracy platform of the German Pirate Party.}

\begin{tabular}{ l r }
\hline
\noalign{\smallskip}
\textbf{Observation period} & 2010/08/13 -- 2013/11/25\\[-0.3ex]
\hline
\noalign{\smallskip}
\textbf{Votes} & 499,009\\[-0.3ex]
\hline
\noalign{\smallskip}
\textbf{Users} & 13,836\\[-0.3ex]
\hline
\noalign{\smallskip}
\textbf{Delegations} & 14,964\\[-0.3ex]
\hline
\noalign{\smallskip}
\textbf{Proposals (\textit{Initiatives})} & 6,517\\[-0.3ex]
\hline
\noalign{\smallskip}
\textbf{\textit{Issues}} & 3,565\\[-0.3ex]
\hline
\noalign{\smallskip}
\textbf{\textit{Areas}} & 22\\[-0.3ex]
\hline
\end{tabular}
\label{tab:dataset}

\end{table}

\section{Description of the Dataset}
\label{sec:dataset}
The German Pirate Party maintains the largest installation of LiquidFeedback with 13,836 registered users, and uses the software to survey the opinion of members.  The German Pirate Party's installation of LiquidFeedback thus represents the largest online community implementing delegative democracy.
In this study we use a complete dataset created from daily database dumps of that installation, ranging from August 13 2010 up to November 25 2013, spanning 1,200 days. The data is available to all party members.

\begin{figure*}[t]
\centering
\centering
 \subfigure[Voting results of \textit{Initiatives}]{
 \hspace{11pt}
\includegraphics[width=0.9\columnwidth]{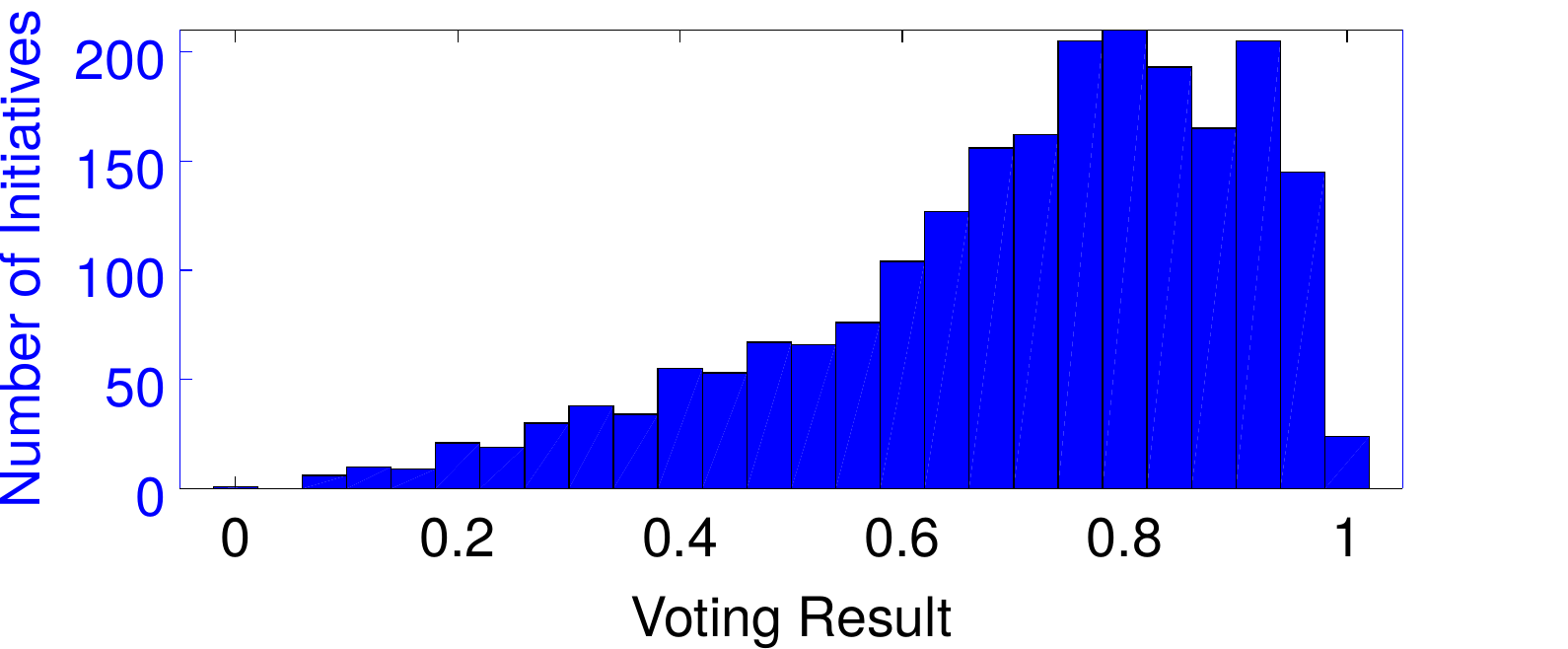}
\hspace{11pt}
 \label{voting_results}
}
 \subfigure[Approval rates per User]{
\includegraphics[width=0.9\columnwidth]{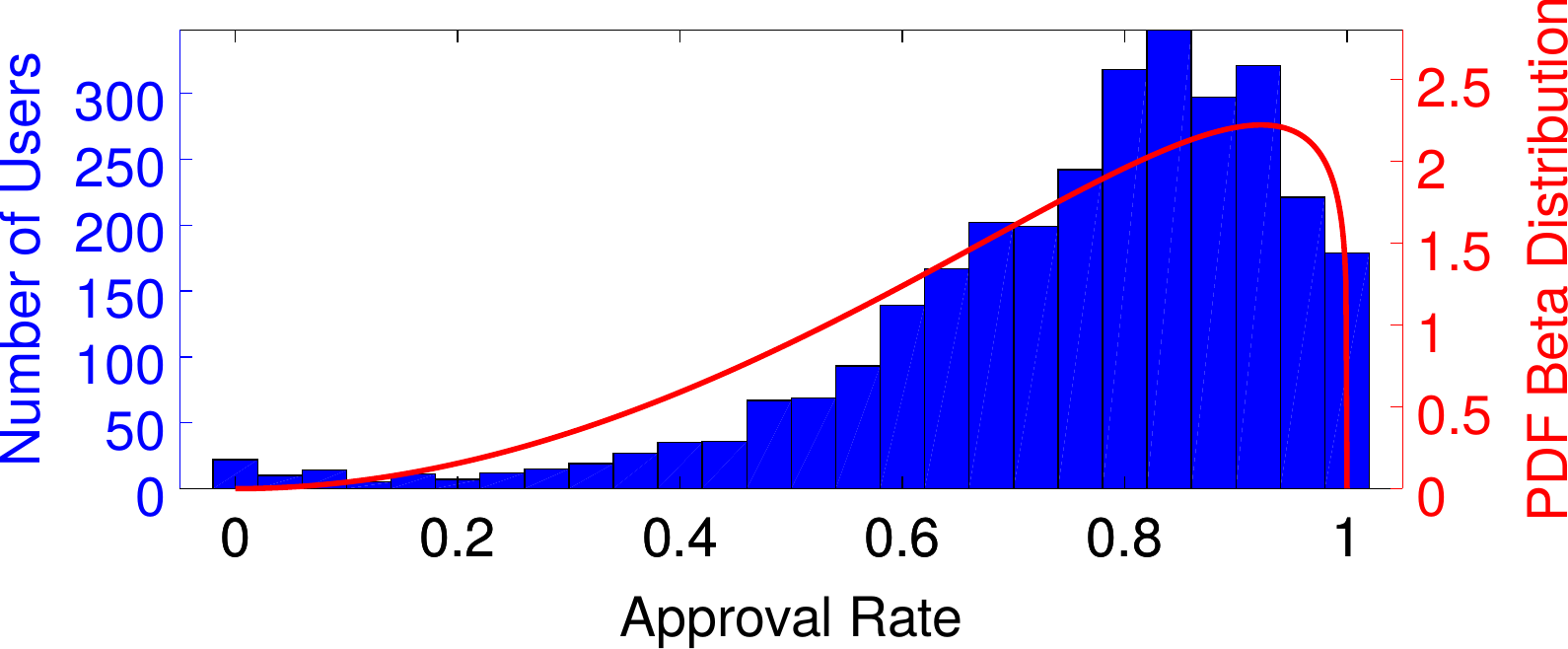}
 \label{agreement_users}
}
\caption{
  \label{fig:support_votes}
  \textbf{Average approval rates per \textit{initiative} and per user.} An approval rate of 1 means maximum approval (all votes have been positive), an approval rate of 0 means minimum approval (all votes have been negative).
  \subref{voting_results}~\textit{Initiatives}. The distribution shows a strong voting bias with a first peak at an approval rate of around 0.75 and a smaller second peak at 0.90.
  \subref{agreement_users}~Users. Histogram of approval rates of users who voted for more than 10 \textit{initiatives}.
We can approximate the per-user approval rate with a beta distribution. Voters show a strong tendency towards approving initiatives with an expected approval rate of 0.71.
}
\end{figure*}

\smallskip
\noindent
\textbf{The LiquidFeedback Platform.}
We give a brief overview over the most important processes and policies within the system and refer to \cite{jabbusch2011liquid,paulin2014through} for a more detailed description.
In LiquidFeedback as used in the German Pirate Party, members can create \textit{initiatives} which are to be voted on to obtain the current opinion of the party members, e.g.\ for collaboratively developing the party program.
Initiatives are grouped into \textit{issues} which group competing initiatives for the same issue.
For instance, if a user proposes an initiative to reduce the emission of $\mathrm{CO}_2$ by subsidising the construction of wind turbines, another user could create a competing initiative to subsidise solar fields.
Furthermore, issues belong to \textit{areas} which represent main topics such as environmental policies.
Each user can create new initiatives, which need a minimum first quorum of supporters for being voted upon. In LiquidFeedback, votes can be delegated to other voters on three levels:
On the global level, meaning that all initiatives can be voted for by the delegate on behalf of the delegating user; on the area level, so that delegations are restricted on an area; or on the issue level.
The actions of every voter are recorded and public, allowing the control of delegates at the expense of non-secret votes.



\smallskip
\noindent
\textbf{Dataset.} In total, the dataset includes 499,009 single votes for 6,517 \textit{initiatives} belonging to 3,565 \textit{issues}. Throughout the four-year observation period, a total of 14,964 delegations where made on the global, \textit{area} or \textit{issue} level, constituting the delegation network (Fig.~\ref{fig:graphdrawing}). The number of active users over the observation period is shown in Fig.~\ref{fig:active_users}. Usage of LiquidFeedback in the German Pirate Party fluctuates with political events in the party. We observe a strong growth in active users after the electoral success of the Berlin Pirate Party in 2011, where 8.9\% of the votes were received. Another point of growth is observed prior to the German federal election in 2012. 180 days after the programmatic federal party congress in 2011, we see a significant drop of active users, when the voting system was used to prepare proposals for the party congress. After the congress, a critical debate on the future role of delegative democracy for the Pirate party started. In a discussion on the effect on \textit{super-voters} -- i.e.\ users with a large share of incoming delegations -- the democratic nature of the system was questioned, and many users 
became inactive.

\section{Voting Behaviour}
\label{sec:voting-behaviour}

In the following, we study different aspects of voting behaviour using the complete voting history in our dataset and the temporal delegation network.

\subsection{Existence and Role of Super-voters}
In order to explore whether super-voters exist, and whether they wield an over-propor\-tional influence in the system,
we calculate the exponent of the power law distributed weight distribution of voters per \textit{issue}, summing over global, \textit{area} and \textit{issue} delegations. The power law exponent is 1.38, indicating that most voters have no delegations and a small set of voters possesses a huge voting weight -- the \textit{super-voters}. There are only 38 voters with more than 100 delegations in the voting history, and we therefore exclude the non-significant statistics for those voters from the figures of this paper.
The practical power of super-voters does not only depend on their voting weight -- it also depends on how often a voter actually participates in votes. One could ask: Are delegates more active than normal users?
We found the overall activity of voters to be power law distributed with an exponent of 1.87 and a median of 8. 3,658 members voted more than 10 times, 1,156 voted more than 100 times and 54 members voted more than 1,000 times. 
The power law exponent of users who received delegations during the observation period is 2.68 with median 64, showing an increased activity and a more homogeneous distribution of activity. 
For controlling this result, we compare it with the exponent of users who delegated their vote at least once to another user. Those users who actively participated in the system have a power law exponent of 2.21 for the number of voted \textit{issues} at a median of 42 -- delegates indeed have a increased activity also when compared to active, delegating users.

To get an insight in the meaning of delegations, we examine the match of voting decisions between delegates and their delegating voters before the delegation. The percentage of votes where both users gave identical ratings (positive/negative) to the same \textit{initiative} is 0.61 whilst any two random voters have an average match of 0.51. As this difference is quite small, we note that delegates do not seem to receive delegations mainly because of shared political views, but they often decide different than their voters in past votes. Delegates in the system then are not expected to represent the opinion of their voters and act independently, giving them a high freedom of action.


Another factor in the power of users are voting results. If votes are narrowly decided, even a small weight gives voters the power to decide votes alone. A histogram for the frequency of voting results is shown in Fig.\ \ref{voting_results}. We see that the distribution is skewed towards positive results with its peak at about $0.8$.
The distribution of support shows a striking similarity to the distribution of ratings in other online communities as described by Kostakos \cite{journals/corr/abs-0909-0237}.
\begin{figure*}
  \centering
    \includegraphics[width=0.70\textwidth]{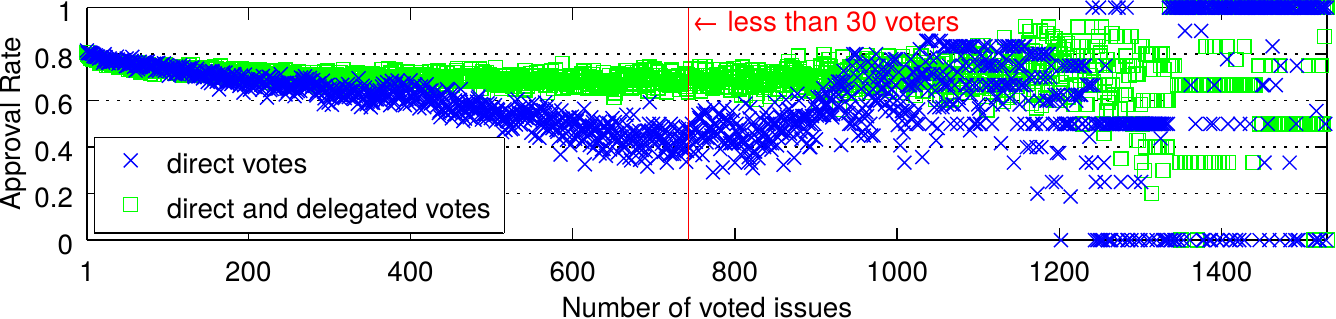}
      \caption{\textit{Approval rate} (Percentage of positive votes) of all voters for the $k$th voted \textit{issue} as a function of k.
  Looking at direct votes only, we see a decrease of approval rates from 0.8 to below 0.5 with higher voting experience, i.e.\ voters become more critical. When including votes made by delegates on behalf of voters, we see only a slight decrease of approval rates and an early stabilisation at about 0.7. Delegates therefore have \emph{a stabilising effect on approval rates}.
  The number of observed votes quickly gets smaller as the number of voted \textit{issues} follows a power law.
   We marked the point where we observe less than 30 direct voters. The increased approval rate of direct votes around the 750th vote therefore indicates the existence of a very small group of active voters with a high approval rate.
   }
   \label{fig:approval_votenumber}
  \end{figure*}
\subsection{User Approval Rates}
In Fig.\ \ref{agreement_users} we plot user approval rates, i.e.\ the percentage of positive votes for each voter.
We exclude users who voted for less than 10 \textit{issues} to ensure significance.
The distribution exhibits a strong bias towards the approval of proposals and reaches the highest numbers at about 0.8 and 0.9.
This distribution closely resembles the overall approval of users for \textit{initiatives}.
Surprisingly, there is a larger number of ``100\%-users'' (in total 160) who voted yes in \textbf{all} of the votes.
We found those users to 
receive a lower number of incoming delegations (1.05 vs. 1.48 on average).
One explanation for this behaviour could be that some users only vote for \textit{initiatives} they support and hope that other \textit{initiatives} won't reach the quorum without their votes.
We can approximate the distribution of user approval by a beta distribution which will prove to be useful later for developing novel power indices. Fig.~\ref{agreement_users} shows a fitted beta distribution as a dashed line. We removed the 100\%-users from the data before learning the parameters to obtain a better fit~\cite{minka00}.

It seems very natural for a democratic voting system without coalitions or party discipline to have a biased distribution of approval rates. As those systems typically include mechanisms to filter out proposals before they reach the voting phase (to prevent an unworkable flood of voting) such as requiring minimum support, the quality of the voted proposals already is relatively high.
Due to selection processes, we argue that most democratic online systems will exhibit a biased distribution of approval rates.

As the approval distribution is close to the \nicefrac{2}{3}~quorum (which typically is required in votes), super voters are expected to have a bigger influence in the voting outcomes.

In order to gain insights in the temporal dynamics of approval rates, we plot the average approval rates for the \textit{k}th vote of all users in Fig.~\ref{fig:approval_votenumber}, illustrating the probability for seeing a positive vote in the first, second etc. vote of a user. Clearly, more experienced users get more critical towards proposals. The learning curve is observed for all users, independent of their activity as measured by the number of voted \textit{issues} -- this e.g. can be seen in the approval rates for users of different activity levels depicted in Fig.~\ref{fig:approval-activity}, which decrease much slower than the learning curve.
The negative impact of the number of votes on the approval rate eventually would lead to a stagnation of the system, as the typical quorum of \nicefrac{2}{3} would be reached by hardly any \textit{initiative}.

\subsection{Impact of Delegations}
Surprisingly, such a stagnation can not be observed in the platform, even in periods when few new users join the system. We plot the effective votes of a user -- i.e.\ all votes including delegated votes made on behalf of a user -- in Fig.~\ref{fig:approval_votenumber} and see that the negative development of approval rates is compensated by delegated votes. 



\begin{figure*}
      \subfigure[\ \ \ \ Approval rate vs. activity]{
      \hspace{0.1\columnwidth}
    \includegraphics[width=0.8\columnwidth]{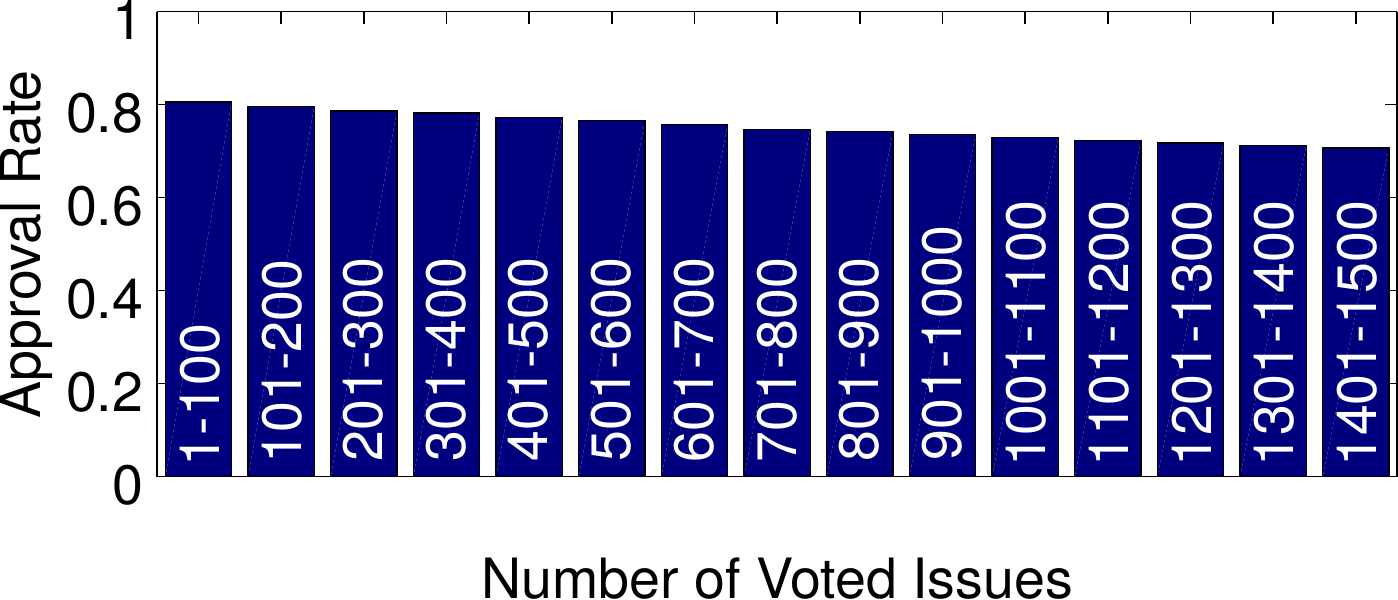}
    \hspace{0.1\columnwidth}
    \label{fig:approval-activity}
  }
      \subfigure[Impact of delegations on approval/agreement rates]{
            \hspace{0.1\columnwidth}
    \includegraphics[width=0.8\columnwidth]{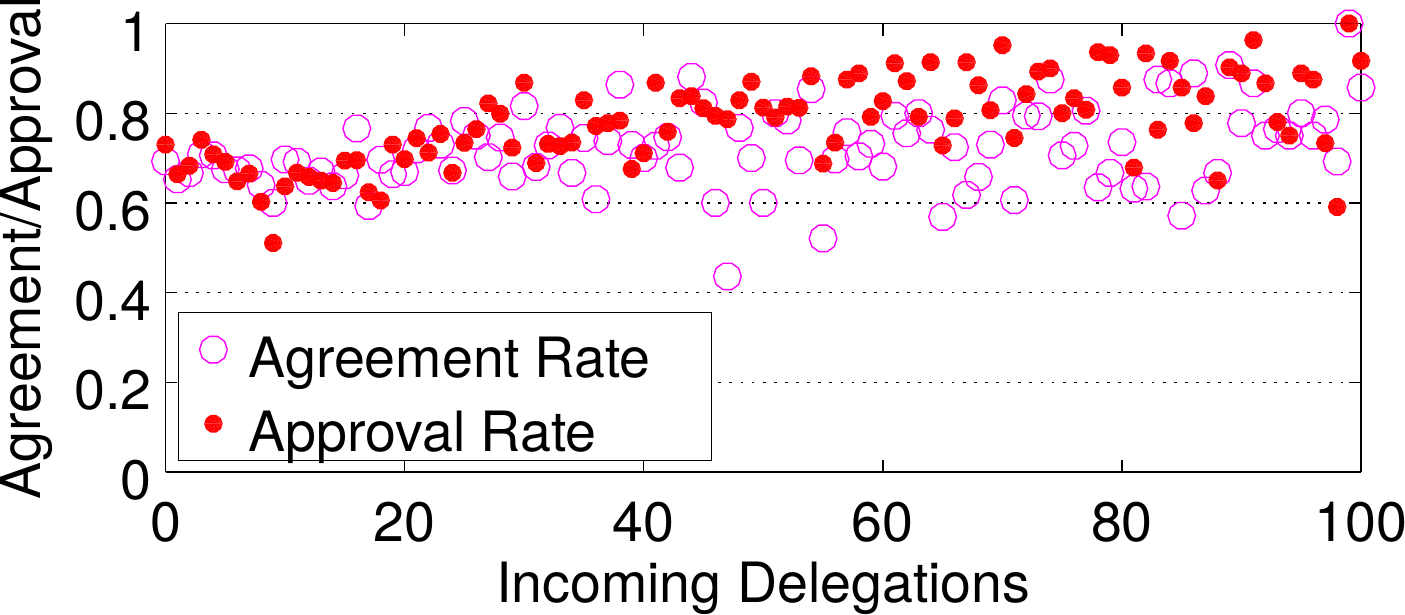}
          \hspace{0.1\columnwidth}
    \label{fig:approval-delegations}
  }
    \caption{\textbf{Voting behaviour.}
    (a)~Active users as measured by the count of voted issues tend to approve \textit{initiatives} less often. The effect is less pronounced than in Fig.~\ref{fig:approval_votenumber}.
    (b)~Approval rate of votes for given weights. Surprisingly, \textit{super-voters} tend to approve more initiatives (approval rate), and tend to agree more often with the majority compared to normal users (agreement rate). Delegations for authors of \textit{initiatives} were ignored to rule out effects of implicit approval.
  }
\end{figure*}

Do these findings imply that \textit{super-voters} are more likely to agree with \textit{initiatives}? And do \textit{super-voters} use their power to turn voting results when voting in favour of initiatives, or do they agree with and vote according to the majority of voters?
Fig.~\ref{fig:approval-delegations} shows the average approval and agreement rate of voters for growing numbers of incoming delegations. The agreement rate is given by the percentage of votes which agree with the majority of voters excluding delegations. We see a positive effect of incoming delegations both on the approval rate and the agreement rate.

In contrast to the intuition that users tend to delegate their votes to users who often vote in favour of proposals, we found no significant differences in the approval rates of users with many delegations in their voting history and normal users. However, as soon as users get many incoming delegations, positive votes get more likely. We hypothesise that voters with many incoming delegations feel social pressure to vote positively and avoid giving a negative vote which would inevitably lead to the failure of a proposal, given the high voting weight. This social control would limit the exercised power of the \textit{super-voters} and stabilise the voting system, effectively preventing political stagnation.



\subsection{Changes Over Time}
\begin{figure}[b!]
  \centering
  \includegraphics[width=1\columnwidth]{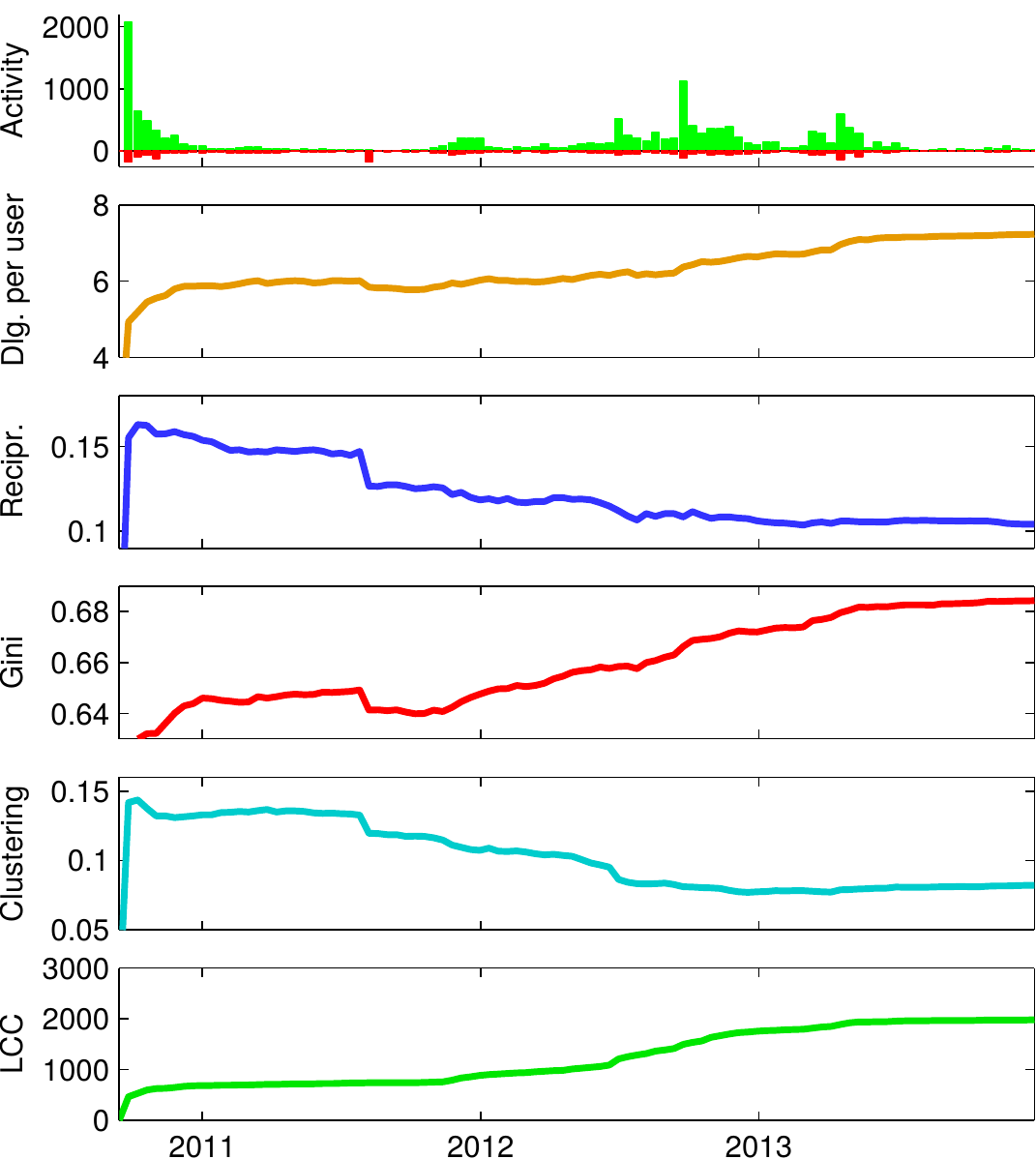}
  \caption{
    \label{fig:histogram}
   \textbf{Changes in the delegation network.} From top to bottom, we show added and removed delegations, changes in the per-user delegation count, inequality of incoming delegations measured by the Gini coefficient and the reciprocity which gives the proportion of mutual delegations. Note that mutual delegations are only permitted for distinct areas or \textit{issues}.
  }
\end{figure}

Since LiquidFeedback is a novel system, its use is still in an emerging stage, and therefore we expect its usage patterns to vary over time. Specifically, we look at the following changes, by analysing the temporal evolution of network-based statistics, shown in Fig.~\ref{fig:histogram}.


\smallskip
\noindent
\textbf{Changes in the distribution of delegations.} While we found the distribution of received delegations to be power law-like,
the inequality of this distribution is not constant, as shown by several statistics in Fig.~\ref{fig:histogram}.  In particular, we computed the Gini coefficient of the delegation network's indegree distribution \cite{kunegis:power-law}, and found that its temporal behaviour is increasing, i.e., the inequality of the number of received delegations increases over time.  This is consistent with a consolidation of the network, i.e., the emergence of \textit{super-voters} and a stronger concentration of power.

\smallskip
\noindent
\textbf{Changes in reciprocity.} We measure the reciprocity of the delegation behaviour as the ratio of delegation edges for which a reciprocal delegation edge exists, to the total number of reciprocity edges, and observe that this value decrease over time.  This would indicate the the community is going away from a set of small groups of voters that delegate to each other, to a community in which most delegation edges go to \textit{super-voters} who do not delegate back.  We must note however that reciprocal delegations are only possible for delegations in different areas, as the set of delegations in a single area must not form cycles.


\smallskip
\noindent
\textbf{Changes in clustering.} We measured the clustering coefficient, i.e., the probability that two neighbours of a voter are themselves connected, within taking into account edge directions \cite{b228}.  This clustering coefficient is decreasing over the lifetime of the network while the largest connected component (LCC) is growing, indicating again that the delegation network is slowly becoming less like a friendship network, and more like a bipartite networks of \textit{super-voters} connected to normal voters.


\section{Voting Power}
\label{sec:measuring_power}

With increasing number of incoming delegations the influence of a delegate increases.
However, the influence on a given \textit{issue} can be very non-linear. As an extreme
example consider an \textit{issue} with voters with delegation counts: $5,4,1$, given a quorum of  \nicefrac{1}{2}.
In this case the actual voting power of the voters with delegation counts $4,1$ is equal, since they have to agree on a position in order to have any
impact at all. We can measure power with theoretical indices and with direct measures on the voting history.

\subsection{Theoretical (Uniform) Power Indices}

The political sciences literature knows several power indices
that measure the influence of individual voters in delegative voting situations
\cite{Straffin1994}. The most common ones are Shapley \cite{Shapley1954}
and Banzhaf \cite{Banzhaf1965}.

Those indices can be best described in terms of (simple weighted)
voting games. A voting game consists of a finite set of voters $i \in
V$ together with weights $w_i \in {1,2, \dots}$ for each voter and a
quorum $q$. A subset of voters $S \subset V$ is called {\em
  coalition}. A coalition $S$ is called {\em winning} if its total
weight is bigger than the quorum:
$\sum_{i \in S} w_i \geq q$.
A voter $i$ in a winning coalition $S$ is called {\em swing voter} if
the coalition $S - \{i\}$ is not winning.

\paragraph*{The Banzhaf power index}
The (unnormalised) {\em Banzhaf power index} of a voter $i$ is defined as
$\hat{\beta}_i = \frac{|W_i|}{2^{n-1}}$,
where $n$ is the number of voters $|V|$, $2^{n-1}$ is the number of coalitions that $i$ is a part of and $|W_i|$ denotes the number of winning coalitions where $i$ is a swing voter. The standard Banzhaf
index $\beta_i$ is the normalization of $\hat{\beta}_i$ that makes all
indices add up to $1$.

\paragraph*{The Shapley power index}
The {\em Shapley power index} measures the number of orderings of
all voters in $V$ where the voter $i$ is ``pivotal''. It is defined as:
\begin{align}
  \phi_i = \sum_{\textnormal{S, $i$ swing for $S$}}  \frac{(|S| - 1)!(n- |S|)!}{n!}.
\end{align}
Both indices can be characterised in probabilistic terms~\cite{Straffin1977}. Indeed,
assume that the vote of a voter $i$ is drawn randomly with probability
$p_i$ for a ``yes'' and $1-p_i$ for ``no''. The {\em individual
  effect} of a voter $i$ is the probability of the voter $i$ making a
difference to the outcome of the entire vote.

Of course, the individual effect will depend on the individual
probabilities $p_i$. Typical assumptions behind existing theoretical power indices are:
\begin{itemize}
  \item Uniformity. Each $p_i$ is chosen from a uniform distribution on $[0,1]$.
  \item Independence. Each $p_i$ is chosen independently.
  \item Homogeneity. All $p_i$ are equal to $p$.
\end{itemize}

\begin{figure*}[t]
  \centering
  \subfigure[Empirical Power]{
    \includegraphics[width=0.65\columnwidth]{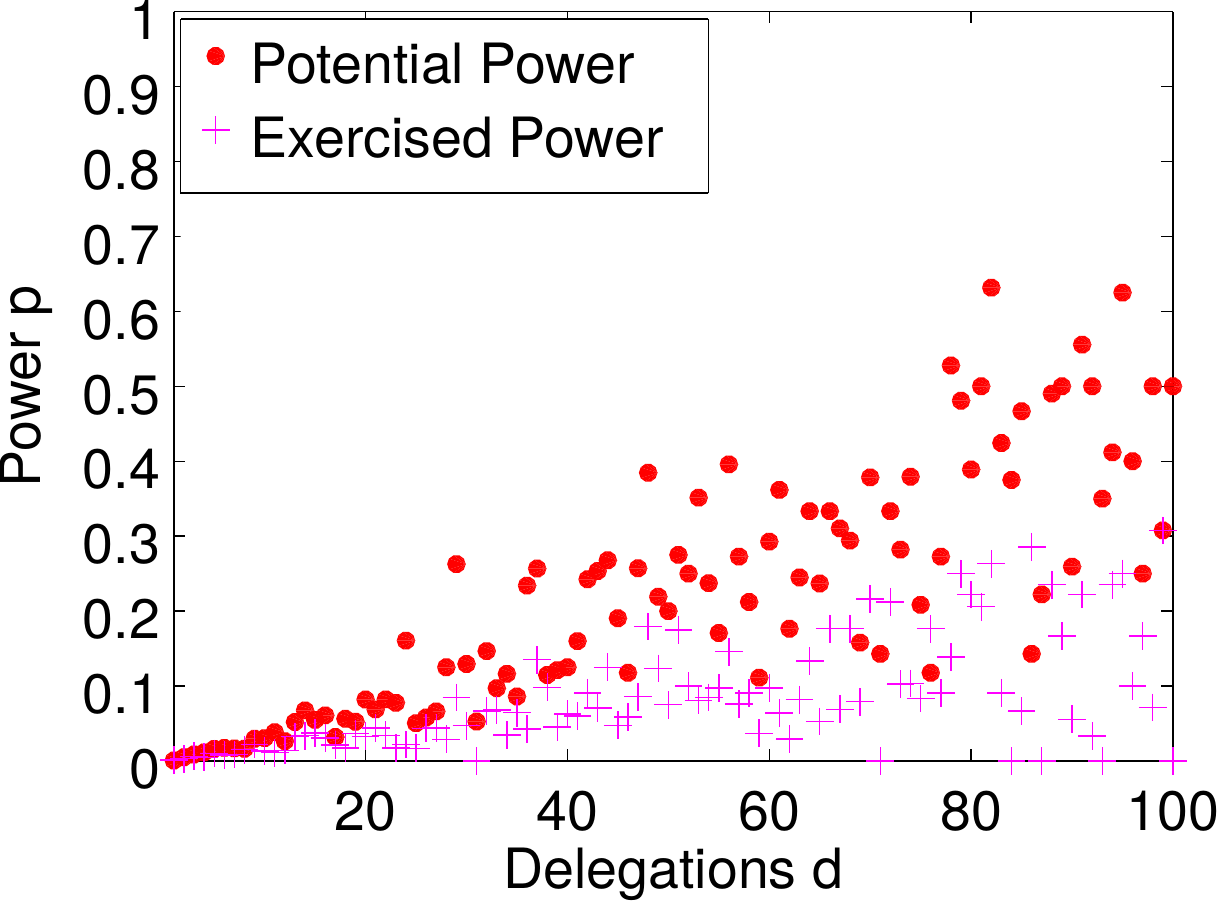}
    \label{fig:empirical_power}
  }
  \subfigure[Theoretical (Uniform) Power Indices]{
    \includegraphics[width=0.65\columnwidth]{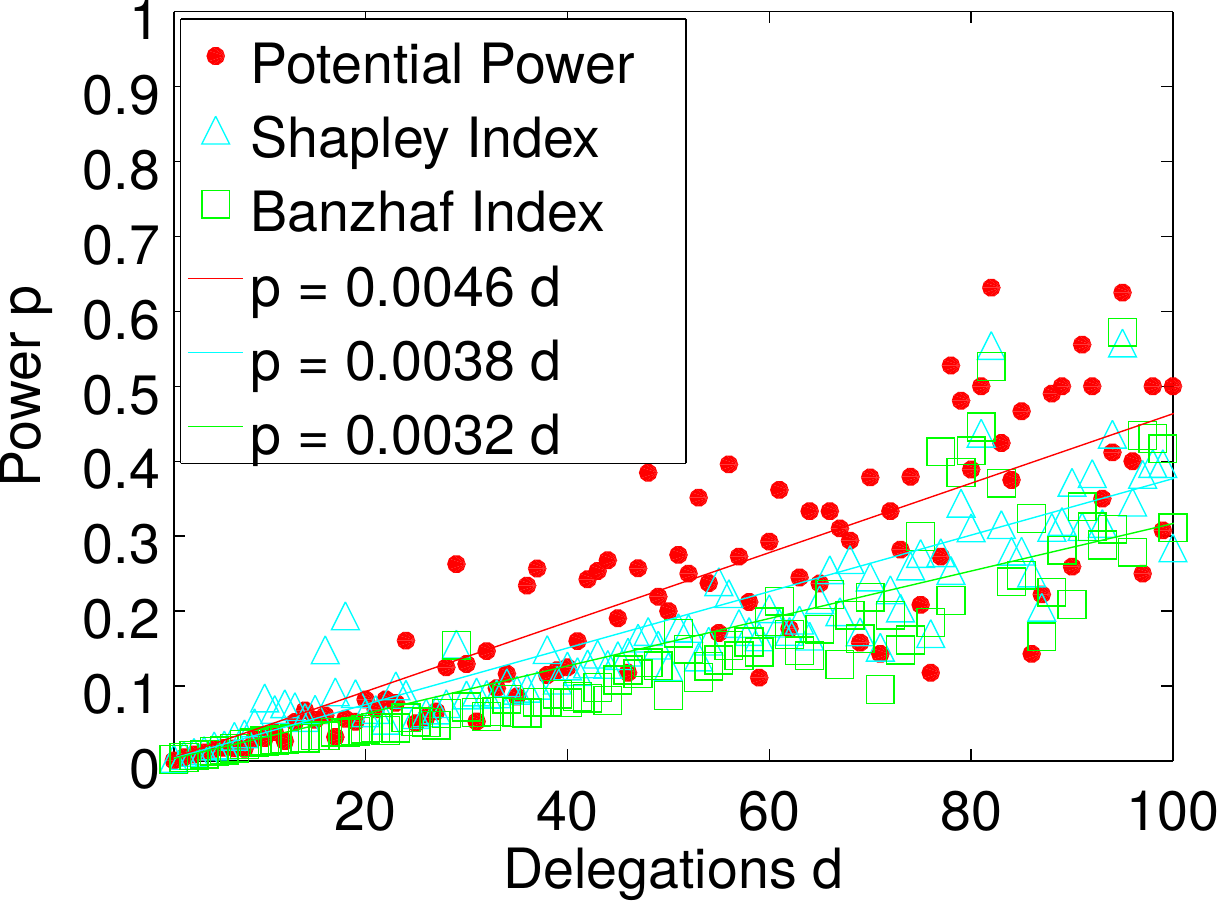}
    \label{fig:applied_power}
  }
  \subfigure[Non-Uniform Power Indices]{
\includegraphics[width=0.65\columnwidth]{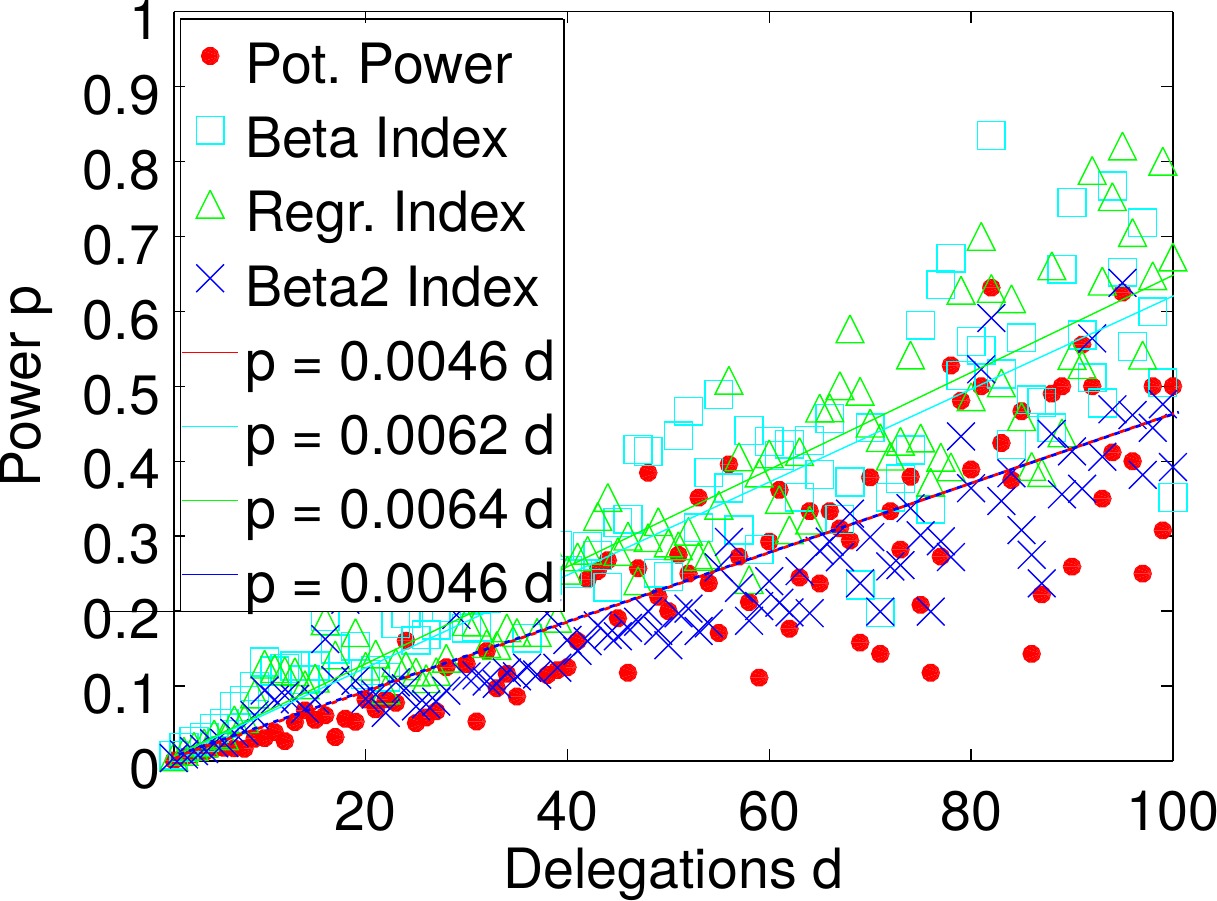}
    \label{fig:nonuniform}
  }
  \caption{Measuring Power of super voters.
    \subref{fig:empirical_power}~Average potential power and exercised power for a given number of delegations.
The exercised power grows significantly slower than the practical and theoretical power, indicating that \textit{super-voters} tend to agree with majority votes.
\subref{fig:applied_power}~Average potential power and average predictions of uniform power indices for given delegation counts. The uniform indices under
-estimate the voting power in the LiquidFeedback data.
    \subref{fig:nonuniform}~Averaged potential power and average power index of the Beta, Regression and Beta2 power index for changing numbers of delegations. The Beta2 index closely predicts the measured potential power.
}
    \label{fig:indexall}

\end{figure*}

It was shown in \cite{Straffin1977} that the Banzhaf index represents the individual
effect of a voter under the assumption of independence and the Shapley
index represents the individual effect of a voter under the homogeneity
assumption. Both indices rely on the uniformity assumption, and most power indices from literature repeat this assumption \cite{Straffin1977,Straffin1994,Gelman.Katz.ea2002MathematicsandStatistics}.


\subsection{Empirical Power}
Theoretical measures of power are based on simulation. With the large number of observations available from the LiquidFeedback dataset, we are able to directly measure power as the ability of a voter to influence voting outcomes.
\paragraph*{Potential Power}
The ability to decide a vote is calculated with the sum of weights of positive $W^p_m$ and negative $W^n_m$ votes in a voting $m$, testing if the weight $w_{im}$ of voter $i$ is bigger than the distance to quorum $q_m$ without $i$:
\begin{align}
\gamma^p_{im} =[
w_{im} >
\underbrace{q_m \cdot \left(W^p_{m} + W^n_{m}\right) - W^p_{m} + w_{im} \cdot v'_{im}}_\text{votes missing to reach quorum without voter i}
> 0
]
\end{align}
where $v'_{im}\in\{0,1\}$ indicates the decision of voter $i$ in voting $m$ and we use Iverson brackets so that $\gamma^p_{im} \in \{0,1\}$.

\paragraph*{Exercised Power}
Similarly, we can look at the actual vote of voters and see whether the power actually was used to \emph{reverse the voting result}:
\begin{align}
\gamma^e_{i} =  \big[(\underbrace{\frac{W^p_m - w_{im}\cdot v'_{im}}{W^p_m + W^n_m - w_{im}} > q_m}_\text{voting result without voter i})
\neq
(\underbrace{\frac{W^p_m}{W^p_m + W^n_m} > q_m}_\text{actual voting result})\big]
\end{align}

Looking at the voting history, the impact of delegates on voting outcomes can be easily estimated by subtracting delegations from vote counts.
Without the delegations, the vast majority of 84,9\% of the results remains unchanged -- only one in six voting outcomes is not identical to the outcome of a hypothetical direct democratic system.

\emph{Does that mean that \textit{super-voters} are not as powerful as they were thought to be?} To answer that question, we calculate the potential and the exercised power in Fig.~\ref{fig:empirical_power} to empirically measure the power of voters. We observe that the ability to decide votes grows approximately linearly with the voting weight. The exercised power measured as the percentage of reversed votes grows significantly slower than the potential power -- \textit{super-voters} use their power relatively less often than ordinary delegates. This explains the positive influence of delegations on the majority agreement observed in Fig.~\ref{fig:approval-delegations}. The average ratio between theoretical and user power is $0.34$ -- powerful users reverse the result of a voting in only one of three votes.
We find a small but significant negative correlation between power and exercised power with $\rho = -0.26$ ($p < 0.05$).

In theory, power indices are supposed to correspond with the potential power of users. To test this, we calculate the Banzhaf and Shapley index for every vote and show the average predicted power in Fig.~\ref{fig:applied_power}. On our data, both theoretical power indices fail to approximate the potential voting power. Instead, the Shapley index and the Banzhaf index understate the potential power of users and predict a growth rate that is lower than what we find.

We are not interested in predicting the exercised power, as our main focus is on the prediction and recognition of high potential power and the danger of power abuse. Though potential power might not be used at a given time, there is no reason to assume that this behaviour is stable.

\subsection{Non-Uniform Power Indices}

The limited alignment of existing power indices with observed voting behaviour suggests that some of the fundamental assumptions behind those indices are not applicable for our data. Existing power indices are based on what we call a uniformity assumption, i.e.\ that users vote with equal probability in favour or against a proposal. Historically, there was no extensive voting data available to test this assumption.
For online platforms such as LiquidFeedback, we have enough data to observe a voting bias~\cite{journals/corr/abs-0909-0237}.
Our findings on the distribution of voting results and user approval rates shown in Fig.~\ref{fig:support_votes} allow us to overcome this over-simplifying assumption of uniformity.

In this section, we \emph{propose generalisations of the Banzhaf and Shapley power index} which allow to model non-uniform distributions of approval rates (as observed in our data).

\paragraph*{Beta index}

The user approval rate $p_i$ approximately follows a beta distribution. Under a beta distribution, this parameter is sampled from
\begin{align}
p_i \sim \frac{1}{B(\alpha,\beta)} p_i^{\alpha-1}(1-p_i)^{\beta-1}.
\end{align}
For parameter estimation, we remove the extreme cases of users with 100\% approval and apply the maximum-likelihood estimate given by Minka~\cite{minka00} to obtain $\alpha = 3.00, \beta = 1.17$.
We display the probability density of the beta distribution in Fig.~\ref{agreement_users};

We first define a generalisation of the Banzhaf index based on the beta distributed $p_i$, the beta power index.
The intuition behind this index is identical to the Banzhaf index: the power of a voter corresponds to the fraction of coalition constellations in which the user is a swing voter.
To create a non-uniform power index, we re-weight the permutation of possible coalitions by their probability under beta-distributed $p_i$ for every voter.
Every voter $i \in V$ has an assigned probability $p_i$ for approving a proposal and users are independent.
For calculating the beta power index $\beta'$, we again calculate all possible coalitions $S \subset V$ and weight them by their probability
\begin{align}
\label{eq:beta-index}
\beta'_i = \int_{0}^1 \cdots \int_{0}^1 \left( \sum_{S \in W} \prod_{j \in V} p_j^{v_{jS}} (1-p_j)^{1-v_{jS}}  \right) \nonumber \\
\prod_{j \in V} Beta(p_j \mid \alpha , \beta)  \ d p_1 \cdots d p_V
\end{align}
where $W$ denotes the set of winning coalitions in which $i$ is a swing voter and $v_{jS}$ is 1 if voter $j \in V$ of coalition $S$ voted ``yes'' and 0 otherwise.
The probability of a coalition is given by a multinomial distribution with success probability $\vec{p} = (p_1,...,p_V)$, the beta distributed approval rates.

Beck \cite{beck1975note} noted that the probability of a tie is very small under such a model -- this finding is trivial and indeed in the whole LiquidFeedback dataset only one \textit{initiative} exhibits a tie.
Gelman et al. claim that models based on binomial distributions with $p \neq 0.5$ would not be useful because of the small standard deviation \cite{Gelman.Katz.ea2002MathematicsandStatistics}.
In our evaluation, we demonstrate that this interpretation is wrong and give a natural explanation by the generative process of the individual approval probability $p_i$ which is sampled from a beta distribution with a possibly large variance.

It is evident that the beta index represents a generalisation of the Banzhaf index -- we can choose symmetric (i.e.\ equal) beta parameters to retain the original index.

\paragraph*{Regression index}

Another observation besides biased user approval rates is the impact of delegations on the approval rate shown in \ref{fig:approval-delegations}.
To model this influence, a logistic regression can be trained to adapt the approval rates for changing weights for an alternative power index.
The regression function is given by
\[p_i = \frac {1}{1+e^{-(\beta_0 + \beta_1 x)}}.\]
We again remove users with 100\% approval rate from our data and learn the regression parameters $\beta_0 = 0.7933$ and $\beta_1 = 0.0036 $.
The regression predicts an approval probability $p_i$ of $0.69$ at a weight of 1 and $0.76$ at a weight of 100.

For obtaining the regression power index $\rho_i$, we weight possible coalitions as the product of all approval rates predicted by logistic regression based on coalitions $W$ where voter $i$ is a swing voter:
\begin{align}
\rho_i = \sum_{S \in W} \prod_{j \in V} p_j^{v_{jS}} (1-p_j)^{1-v_{jS}}
\end{align}

\paragraph*{Beta2 index}
The assumption of independence made by the Banzhaf index implies that voters have inhomogeneous opinions and that there is frequent disagreement in votings, i.e.\ there exist opposing factions within the party.
In contrast, the Shapley index assumes that all voters share a similar opinion on a particular \textit{initiative} and therefore agree with it with the same probability $p_i = p, \ \forall i \in V$. However, $p_i$ in the Shapley index is sampled from a uniform distribution.

We modify the index by sampling $p$ from the same beta distribution employed for the beta index: \mbox{$p \sim Beta(\alpha, \beta)$}
with  $\alpha = 3.00, \beta = 1.17$.
This index assumes that voters share a homogeneous opinion on \textit{initiatives}, and that there is a positive voting bias to accept proposals.
For the overall calculation of the beta2 power index $\beta''_i$ we sum over possible coalitions $S \subset V$, weighted by their probability:
\begin{align}
  \beta''_i = \int_{0}^1 \left[ \sum_{S \in W} \prod_{j \in V} p^{v_{jS}} (1-p)^{1-v_{jS}} \right] Beta(p \mid \alpha, \beta) \ dp
\end{align}
where $W$ again denotes the set of winning coalitions in which $i$ is a swing voter,  $v_{jS}$ the approval of voter $j \in V$ in coalition $S$.

\begin{table}[b]
\center
\caption{
  \textbf{Performance of power indices.} Perplexity and squared prediction error for the uniform power indices by Banzhaf and Shapley and the non-uniform power indices presented in this paper, evaluated on the complete voting history of the LiquidFeedback system. Lower perplexity values indicate a better model fit. The Beta2 index proposed earlier outperforms existing and other competing power indices.
}
\begin{tabular}{ l | c | c }
\hline
\noalign{\smallskip}
\textbf{Model} & \textbf{Squared Error} & \textbf{Perplexity}  \\
\hline
\noalign{\smallskip}
Shapley\cite{Shapley1954} & 0.903 & 78.6 \\
Banzhaf\cite{Banzhaf1965} & 1.320 &297.9 \\
Beta power index &2.220 & 227.8 \\
Regression power index  &2.266 &  232.0 \\
Beta2 power index & \textbf{0.627} & \textbf{76.6} \\
\hline
\end{tabular}
\label{tab:evaluation}
\end{table}
\subsection{Evaluation}
The potential power, measured in the same voting system over thousands of votings for voters with changing voting weights, enables an objective evaluation of the predictive performance of power indices.
We compare the presented uniform power indices with the proposed non-uniform inidices by their predictive performance on the potential power of voters.
To obtain the index values, we use Monte-Carlo simulation, first randomly sampling approval rates and subsequently sampling individual votes. We performed $1,000,000$ runs for each voting. The evaluation was run on a standard desktop computer.

\smallskip
\noindent
\textbf{Graphical evaluation.} 
Fig.~\ref{fig:nonuniform} compares the potential power of voters with the prediction of the proposed non-uniform power indices. For every \textit{initiative} in the voting history, power indices are computed based on the voting weights. The resulting indices then are averaged for each voting weight.
We observe that both the beta and the regression index over-estimate the power of users. The regression index predicts values slightly higher than the beta index.
In contrast, the beta2 index predicts values very close to the true potential power and closely resembles the gradient of the measured power, giving a good assessment of the influence of \textit{super-voters}.

\smallskip
\noindent
\textbf{Quantitative evaluation.} 
For a quantitative comparison of the power indices, we evaluate the prediction both on the global and on the local level.
On the global level, we try to predict the average power of \textit{super-voters} as in  Fig.~\ref{fig:indexall}. We measure the closeness of the prediction as the sum of squared errors of the predicted theoretical power and the measured potential power for voting weights $w_i \in [1,100]$.
The results are shown in Table \ref{tab:evaluation}. The biggest deviations are found for the regression, beta and Banzhaf index, indicating that the independence assumption is violated in the voting system. For the Shapley index, we get a significantly lower value and the beta2 index provides the closest approximation.

On the local level, we make use of the extensive voting history to compare the observed potential power of voters -- the ability to decide a vote -- to the predicted power index of every user.
Following the probabilistic interpretation of power indices~\cite{Straffin1977}, a power index corresponds with the predicted probability of a voter having potential power. We computed this probability for every voter in each vote. Now, given the measured potential power of a voter, we can calculate the log-likelihood of the observed power in the voting history. Formally:
\begin{align}
\log L = \sum_{m=1}^M \sum_{i\in V_n} \log(p(\gamma^p_{im}))
\end{align}
where $M$ is the number of \textit{initiatives}, $V_n$ is the set of voters participating in the vote over \textit{initiative} $m$ and $\gamma^p_{im}$ indicates the potential power of voter $i$ in voting $m$.

The likelihood can then be used to calculate the perplexity, a common measure for the predictive quality of a probabilistic model.
The perplexity is defined as
\begin{align}
\mathit{perplexity} =  2^{- \frac{1}{M} \sum_{i=1}^M \log L}
\end{align}

Following the perplexity scores, the beta2 index outperforms all other indices. The Shapley index yields the second best result. The beta index is slightly better than the regression index and the Banzhaf index performs worst.

\section{Discussion}
\label{sec:discussion}

The observed performance of the indices allows us to evaluate the assumptions behind these models. First, we note that the indices based on the independence assumption of voters perform significantly worse than the indices based on the homogeneity assumption, implying that voters share a common opinion given by the approval rate of a vote. We found that the integration of the observed positive influence of delegations on the approval rate by the regression index leads to worse performance. The effect seems to be more complex and has to be examined in future work. Modelling voters homogeneously -- e.g.\ sampling the approval rate independent of the voting weight -- yields significantly better results.

Including observed voting bias in power indices leads to an overall better predictive quality of both indices, measured by lower perplexity. However, only for the homogeneous indices we observed a better global prediction. Independent approval rates sampled from a uniform distribution better approximate homogeneous voting behaviour than biased approval rates.

The proposed beta2 index, a biased generalisation of the Shapley index, gives a precise prediction on the overall power distribution in a voting system with delegations. We can predict the ability of delegates to decide votes by sampling sets of voters and calculating the beta2 index. The beta distribution parameters can be learned from voting history or taken over from similar voting platforms. With those predictions, qualified statements about the distribution of power in voting systems can be made and discussions objectified.

Both the analysis of voting behaviour and the empirical measurement of potential and exercised power exhibit a responsible exercise of power by \textit{super-voters}. We believe that this is due to a responsible selection of delegates, the social control in an enforced public voting and the risk of the immediate loss of voting power by recall of delegations.

\section{Conclusions}
\label{sec:conclusion}

Platforms for online delegative democracy are likely to gain relevance for political movements and parties in the future. Understanding the voting behaviour and emergence of power in such movements represents an important but open scientific and pressing practical challenge.
In this paper we have studied (i) \emph{how people vote} in online delegative democracy platforms such as LiquidFeedback, and \emph{how they delegate votes} to what we call \textit{super-voters}.
This has motivated us to (ii) better \emph{understand the power they have over voting processes}. In particular, we explored (iii) \emph{the theoretical, potential as well as the exercised power} of super voters in online delegative democracy platforms. Towards that end, we employed the Banzhaf and Shapley power index but found conflicts between the assumption of uniformity of voting behaviour made by both indices and the observed voting bias. We have thus introduced and evaluated a new class of power indices that (a) generalises previous work based on beta distributed voter approval and (b) achieves significantly better predictions of potential voting power in our evaluation. To the best of our knowledge, our evaluation based on a large voting history represents an innovative objective evaluation of power indices.

Our work illuminates the potential of online delegative democracy platforms and sheds light on the power of \textit{super-voters} in such systems. While we find that the theoretical and potential power of \textit{super-voters} is indeed high, we also observe that they stabilise the voting system and prevent stagnation while they use their power wisely. \textit{Super-voters} do not fully act on their power to change the outcome of votes, and they vote in favour of proposals with the majority of voters in many cases. This suggests that potential limitations of online delegative democracy platforms (such as the domination of \textit{super-voters} over regular voters) can be -- and indeed are -- alleviated by the behaviour of super voters in such systems to a certain extent.


\smallskip

\paragraph{Acknowledgments}
We thank Claudia Wagner and Lena Hegerfeld
 for their early contribution of ideas to this paper. 
This work was supported by the research network \emph{Communication, Media and Politics} (KoMePol) at the University Koblenz--Landau.

%
%
%
%
%

\let\oldbibliography\thebibliography
\renewcommand{\thebibliography}[1]{%
  \oldbibliography{#1}%
  \setlength{\itemsep}{-0.1pt}%
}
\bibliographystyle{aaai}
\bibliography{lqfb}

\end{document}